# DIELECTRIC PROPERTIES OF FLASH SPARK PLASMA SINTERED $BaTiO_3$ AND $CaCu_3Ti_4O_{12}$

Charles Manière[a*], Guillaume Riquet[a], Sylvain Marinel[a]

(a) Normandie Univ, ENSICAEN, UNICAEN, CNRS, CRISMAT, 14000, Caen, France

**Keywords**

Flash Spark Plasma Sintering; Barium Titanate; $CaCu_3Ti_4O_{12}$; Dielectric Constant; Phase Decomposition

**Abstract**

Flash sintering is an approach allowing reducing the sintering time to mere seconds. To improve the microstructures and the properties of flash sintered specimens, this process has been successfully adapted to pressure assisted sintering such as the spark plasma sintering. This work is the exploration of the potential of this ultra-rapid sintering process for the improvement of the dielectric properties of well-known materials such as $BaTiO_3$ and $CaCu_3Ti_4O_{12}$. In particular, we focus on the potential improvement of the dielectric loss, colossal permittivity and microstructures of these materials after ~20 sec sintering and annealing.

---

* Corresponding author: **CM**: Laboratoire de cristallographie et sciences des matériaux (CRISMAT), 6 Bvd du maréchal Juin 14050 CAEN CEDEX 4, France
Ph.: +33.2.31.45.13.69 ; *E-mail address*: charles.maniere@ensicaen.fr

Dielectric materials such as $BaTiO_3$ and $CaCu_3Ti_4O_{12}$ are of high interest for the improvement of capacitors, resonators, supercapacitors, etc[1–5]. In this field, the research is oriented in the miniaturization of the electronic components by keeping microstructures at nanoscale while maintaining high performances [6–8]. Sintering is a key step for electroceramic materials and has a high impact on the microstructure[9,10], the crystal defects, the stoichiometry, and the dielectric properties[11,12]. In particular, Spark Plasma Sintering (SPS) has demonstrated a strong potential to generate colossal dielectric properties[11,13]. These properties are often associated with the compound reduction phenomenon [14] possibly combined with an internal barrier layer capacitor behavior (IBLC)[15], electrode effects and/or hopping conduction mechanism[11,16,17]. The IBLC ideal model consists of polycrystalline materials with thin electrically insulating grain boundaries and semiconducting large grains. It is generally assumed that the SPS highly reductive atmosphere induces oxygen vacancies playing a key role in the semiconducting behavior in both $BaTiO_3$[16] and $CaCu_3Ti_4O_{12}$[14] oxides. The hopping is another mechanism which may occur in mixed valence semiconductor materials by charge hopping from one site to another, therefore impacting [11,16] the dielectric properties. However, the compound reduction occurring in SPS has a very detrimental impact on the dielectric losses, which are generally very high for colossal permittivity materials. It is often observed that long SPS cycle at high temperatures generates colossal permittivity with high dielectric loss[13]. A subsequent annealing in air is usually required to decrease the dielectric losses; however, it is generally accompanied by a significant reduction of the permittivity[11]. Controlling the sintering process is therefore of primary importance to attain the objective of high permittivity and low dielectric losses. For this goal, different strategies can be implemented. For instance, some groups [[18–20] have designed core-shell composite structure, where semiconducting grains are surrounded by silica, prior sintering. This composite structure was shown to be efficient for decreasing the dielectric losses.

In this work, our strategy is to reduce the sintering time down to a few seconds to partially decrease the compound reduction phenomenon while using the abrupt and high thermal profile for promoting "flash grain growth"[21]. In this way, we can expect to gather the optimal conditions of the IBLC theory[22], i.e., large semiconducting grains and fine electrically insulated grain boundaries. We also expect to obtain a balanced semiconducting behavior by limiting the reduction phenomena which generally ends-up to high dielectric losses. In sum, an ultra-rapid process is explored to find a window between excessive reduction of SPS (high permittivity, high loss) and conventional ways (moderate permittivity, low loss).

Since the discovery of flash sintering[23–25], this process has been applied to various approaches such as microwave sintering[26,27] or SPS[28,29]. The last approach is very promising as it allows combining high electrical current and high pressures[30]. By an electric current concentration[31], it has been shown that this "flashSPS" approach can be extended to nearly all materials[32] (from dielectric to metals). In this work, a similar flashSPS configuration is used for the ultra-rapid sintering of $BaTiO_3$ and $CaCu_3Ti_4O_{12}$. All sintering experiments were performed on SPS device model "FCT HPD25" using a punch and die set for the fabrication of pellets of 15 mm diameter and 2 mm height. The $BaTiO_3$ were purchased from Chempur (99.6 % pure, 85-128 nm) and $CaCu_3Ti_4O_{12}$ powder were conventionally synthesized[33] using the following powders, $CaCO_3$ (99,95% Alfa Aesar), CuO (99,7% Alfa Aesar) and anatase-$TiO_2$ (99,9% Materion). A comparison of microstructure and dielectric properties is investigated using a 100 K/min usual heating rate and in flash condition (~ 1000 K/min). The flashSPS configurations are the same as for the ref[32]: the electrical current is concentrated around the powder bed by electrically insulating the die (via a boron nitride coating on the inner die surface). Compared to ref[32], the flash process is PID controlled by imposing a heating rate of 1000 K/min on the punch temperature. The punch temperature is different from the powder temperature and this difference may increase for high heating rates due to the electric and thermal contact resistance at the interfaces[34–36]. The measured temperature is used for the regulation of the process, but it is not relevant for indicating the sintering conditions of the powder. The sintering advancement is controlled by the displacement curves and for all experiments (SPS and flashSPS), the heating ramp is stopped when the displacement curve indicates the ending stage plateau. For $CaCu_3Ti_4O_{12}$ in classical SPS configuration, a dwell at 950 °C is added to prevent the phenomenon of phase decomposition (see further). For flashSPS, the sintering time is about 20 sec. The dielectric properties were measured from 5 to 1000 kHz, with FLUKE PM6306 RCL meter. The X-ray diffraction diagrams were recorded on XRD Theta-Theta PANalytical'pert Pro diffractometer with CuKα λ = 1.54059 Å. The Scanning Electron Microscopy and EDX map were collected on "SEM Zeiss EVO MA 15".

For both $BaTiO_3$ and $CaCu_3Ti_4O_{12}$, all the sintered pellets were black. For removing the carbon contamination and re-oxidizing the compounds, an annealing in air (150 K/h heating up to 800 °C, holding of 4 hours and cooling at 150 K/h to room temperature) was performed for both materials. After this annealing, the $CaCu_3Ti_4O_{12}$ pellets are still black and $BaTiO_3$ pellets become gray for conventional SPS and white for flashSPS (see figure 1a). The black color was expected for $CaCu_3Ti_4O_{12}$[33], since this oxide is naturally black. For $BaTiO_3$ SPS sintered, the long sintering duration implies a quite significant compound reduction[11]. As a result, a thermal annealing at 800 °C for 4 hours is not long enough to fully re-oxidize the material (it remains gray). On the contrary, the flashSPS sample (sintered in 20 sec) becomes white after annealing, clearly showing that the reduction mechanism was limited. The X-ray diffraction of $BaTiO_3$ is reported in figure 2b, for SPS, flashSPS specimens before and after 800 °C annealing, $BaTiO_3$ SPS sintered, XRD pattern peaks can be indexed to $BaTiO_3$ phase. It is also remarkably observed the broadening of the (200) and other related peaks (220), (103) etc, which can be interpreted by the presence of both a cubic and a tetragonal perovskite phase. This phenomenon has been already observed and discussed in ref[16]. On the contrary, for the 'flashSPS' sintered materials, the corresponding XRD peaks are sharps with the splitting of the (002)/(200) peaks and other related peaks, showing only the presence of the tetragonal $BaTiO_3$ phase. This is consistent with the fact that flashSPS process is beneficial for limiting the material reduction. Concerning the XRD diagrams of $CaCu_3Ti_4O_{12}$ (figure 1c), the difference between the classical SPS and flashSPS suggests a higher temperature for the sample obtained by flashSPS. The XRD pattern of the classical SPS sintered samples only has the peaks corresponding to $CaCu_3Ti_4O_{12}$. For the flashSPS sample, four additional phases appear, namely $TiO_2$, $CaTiO_3$ and $Cu_2O$ (which is fully converted to CuO after annealing at 800 °C). Because of the high temperatures of flashSPS, $CaCu_3Ti_4O_{12}$ phase partially decomposes during this flash thermal treatment.

The polished microstructures of all samples are reported in figure 2. For classical SPS $BaTiO_3$ sintered sample, the grain size is submicronic and the amount of porosity is about 5 %. FlashSPS sintered material has larger grains (~10 μm) with a higher porosity level. In a previous study [32], it was shown by the finite element simulation that a significant temperature gradient develops through the flashSPS sintered sample (for dielectric material), with the highest temperatures on the edges. Experimentally, it is observed that the grain size is significantly larger in the surface than in the bulk of the sample, which is consistent with the modeling result (temperature gradient). Concerning $CaCu_3Ti_4O_{12}$, the grains and porosity are similar between classical SPS and flashSPS configurations. The main difference is the $CaCu_3Ti_4O_{12}$ phase decomposition (for flash sintered material), as revealed by the XRD pattern. In figure 2 (for flashSPS), a white phase appears in the edge of the specimen and 'dark phases' are also observed in some localized areas. The EDX map of this black phase reveals a titanium rich zone (possibly $TiO_2$ observed in XRD) and the white phase is copper rich (possibly CuO observed in XRD). This suggests an overheating on the edges of the specimen as it was previously shown by modeling on dielectric flashSPS [32].

The dielectric measurements (5-1000 kHz) are reported in figure 3. In the following, for simplicity, the real part of the permittivity will be called "permittivity". $CaCu_3Ti_4O_{12}$ has a permittivity between $10^3$-$10^4$ and a loss tangent between 0.1-1. After the 800 °C annealing, the permittivity is almost twice and the loss tangent corresponding to flashSPS is interestingly reduced to a stable value close to 0.1. Concerning $BaTiO_3$ and before annealing, a clear conductive/semiconductive behavior is observed with a very high loss tangent ($\tan\delta > 1$-$10^2$) and a high permittivity (for flashSPS). After annealing, the "SPS" experiment has a high permittivity and a loss tangent closed to 1 (which is still highly dissipative). In comparison, the flashSPS experiment succeeds to preserve the classical values obtained for $BaTiO_3$ by conventional sintering with a permittivity closed to 2000 and a loss tangent below 0.1. Among all experiments, the optimal property is obtained for flashSPS of $CaCu_3Ti_4O_{12}$ after annealing with ~ 4000 of permittivity and a loss tangent of 0.1.

To conclude, we have investigated the impact of flashSPS on the properties of classical dielectric materials such as $BaTiO_3$ and $CaCu_3Ti_4O_{12}$. In general, the SPS reductive atmosphere tends to generate colossal permittivity but the dielectric nature of the materials is annihilated in favor of a conductive/semiconductive behavior. This phenomenon is highly active for $BaTiO_3$ with long sintering time. The significant reduction of the sintering times of flashSPS (20 sec) has a positive impact on this phenomenon while allowing preserving the tetragonal phase of $BaTiO_3$. After annealing it is even possible to reproduce the classical dielectric behavior of $BaTiO_3$ by "flash" approach. For $CaCu_3Ti_4O_{12}$, the reduction phenomenon is less intense than $BaTiO_3$ and a phenomenon of phase decomposition is also observed. The annealing (800°C, 4 hours) helps decrease the flashSPS dielectric losses while increase the permittivity to 4000. The potential role of the $TiO_2$, CuO phases *in situ* generated by the "flash" process should be investigated by a modeling approach. To go further, the impact of flashSPS/annealing, time, temperatures should be finely explored in order to decrease the dielectric loss while more significantly increasing the permittivity *via* IBLC, local hopping mechanisms, etc. FlashSPS is an original opening to escape the classical trend in SPS where colossal dielectric properties are accompanied by high dielectric losses. In particular, the capacity of flashSPS to preserve the powder phase like for BaTiO$_3$ represents an interesting prospect for composites and core-shell structures.

**Acknowledgements**

The help and support of Christelle Bilot and Jérôme Lecourt is gratefully acknowledged.

The authors thank the French Ministry of Research for the financial support of Guillaume Riquet's PhD.

# Figure captions

Fig. 1: a) Sintered pellets after annealing at 800 °C (all pellets were black before annealing), XRD of the flash and conventional SPS before and after annealing for b) BaTiO$_3$ and c) CCTO; the other phases creating during the sintering process are indexed using colored symbols (see lower legend).

Fig. 2: Polished SEM images of the microstructures in the center and the edge of the pellets; below is an EDX map of the localized needle microstructure observe in the edge of flash sintered CCTO pellets.

Fig. 3: Dielectric properties of BaTiO$_3$ and CCTO obtained by conventional SPS and flashSPS, a) before and (b) after 800 °C annealing.

# Figures

Fig. 1: a) Sintered pellets after annealing at 800 °C (all pellets were black before annealing), XRD of the flash and conventional SPS before and after annealing for b) BaTiO$_3$ and c) CCTO; the other phases creating during the sintering process are indexed using colored symbols (see lower legend).

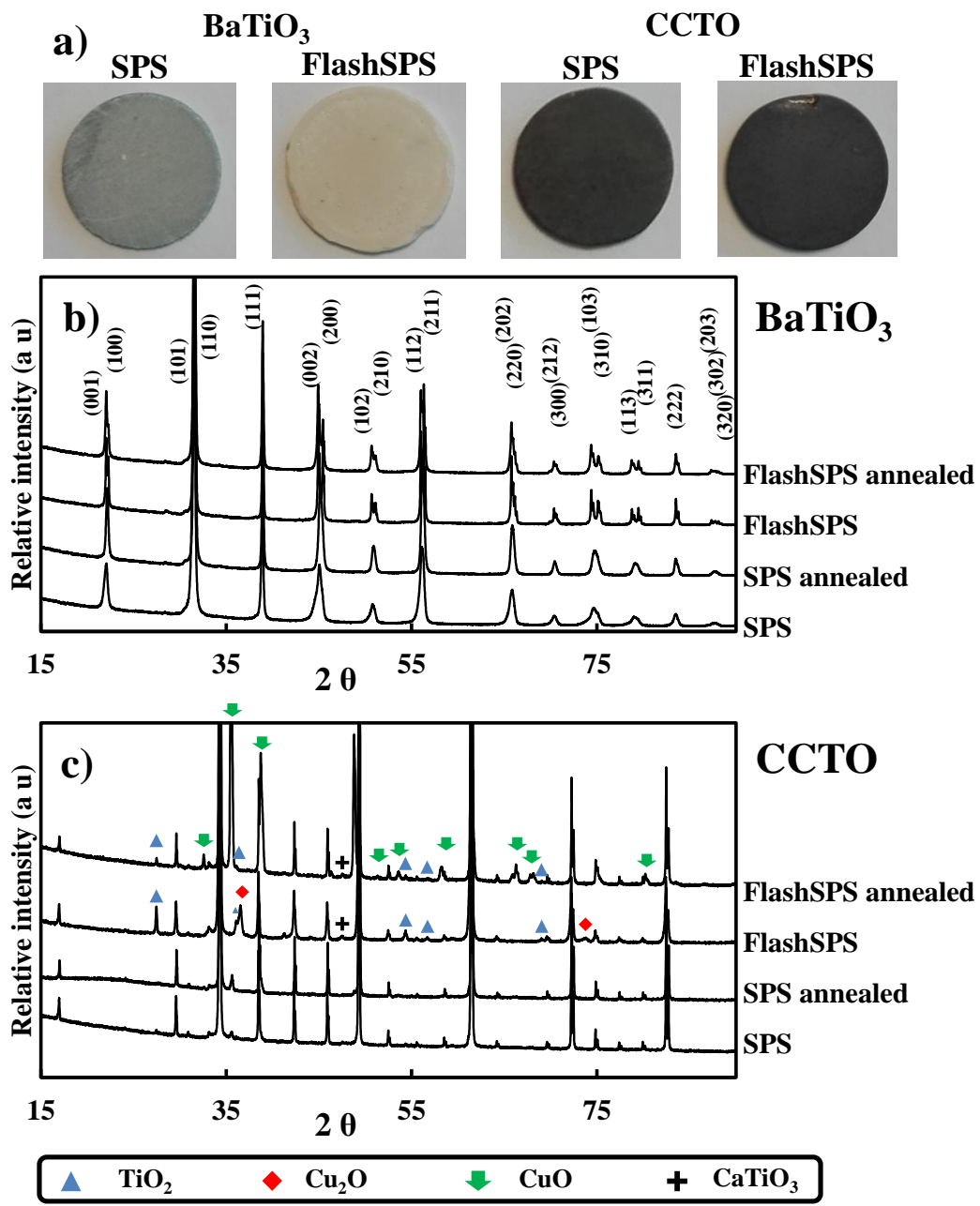

Fig. 2: Polished SEM images of the microstructures in the center and the edge of the pellets; below is an EDX map of the localized needle microstructure observe in the edge of flash sintered CCTO pellets.

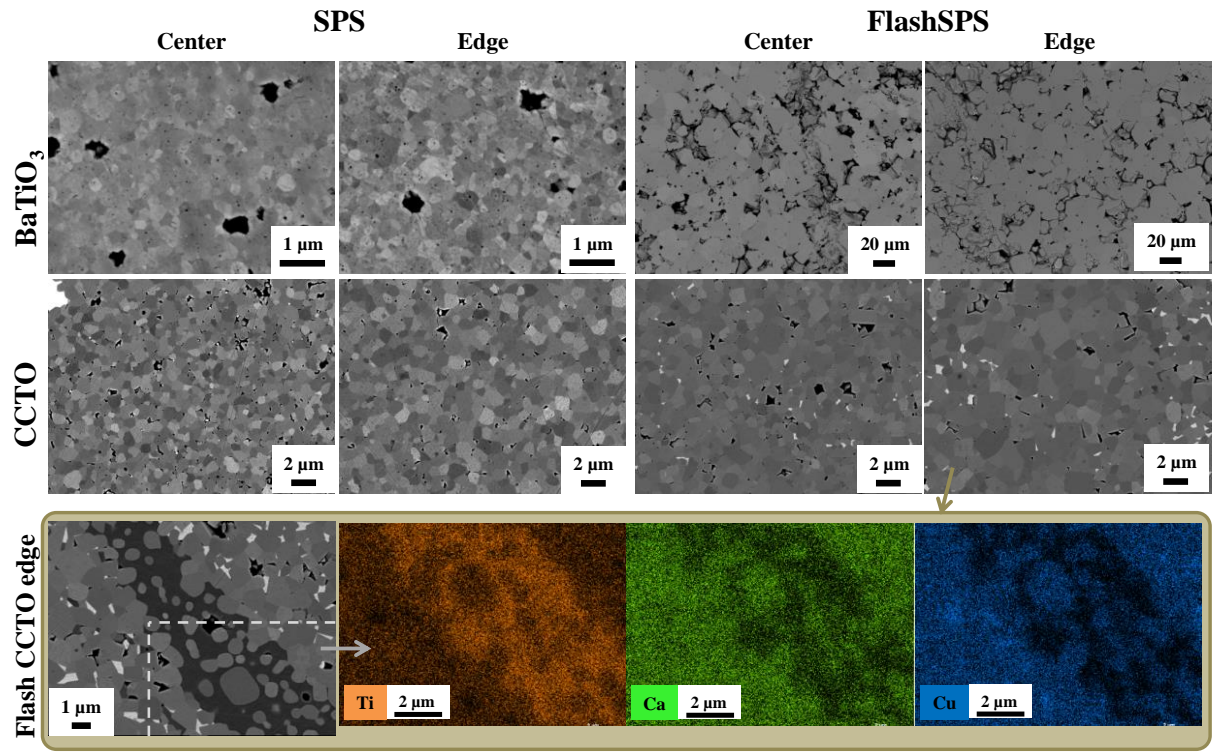

Fig. 3: Dielectric properties of BaTiO$_3$ and CCTO obtained by conventional SPS and flashSPS, a) before and (b) after 800 °C annealing.

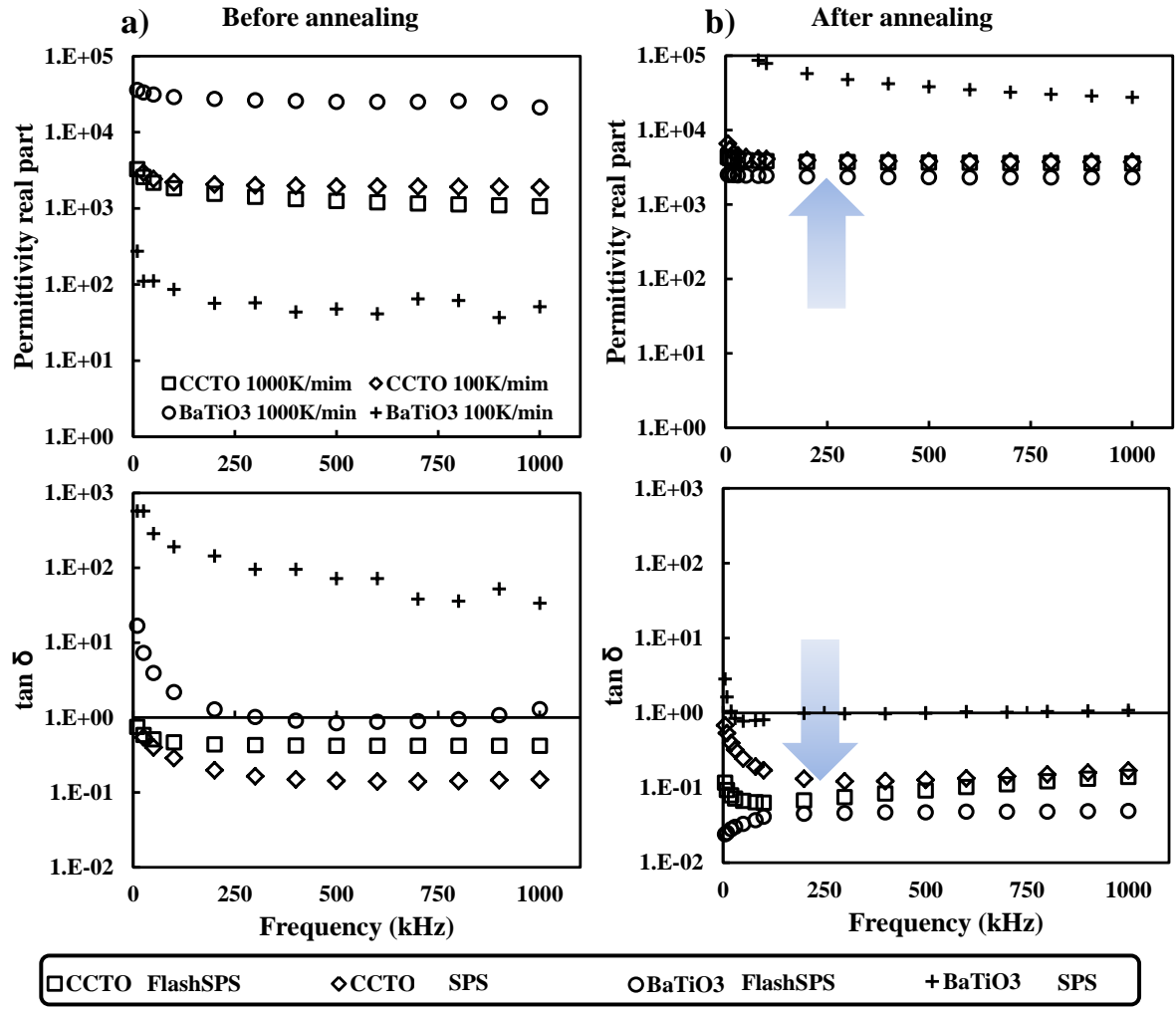